\documentclass[letterpaper, 10 pt, conference]{ieeeconf}  

\usepackage{blindtext}
\usepackage{textcomp}
\usepackage{graphicx}
\usepackage{times}
\usepackage{helvet}
\usepackage{courier}
\usepackage{mathrsfs}
\usepackage{subfigure}
\usepackage{amsmath}
\usepackage{amssymb}
\usepackage{enumerate}
\usepackage{epsfig}
\usepackage{cite}
\usepackage{epstopdf}
\usepackage{algorithm}
\usepackage{algorithmic}
\usepackage{clrscode}
\usepackage{pifont}
\usepackage{amssymb}
\usepackage{diagbox}
\usepackage{multirow}
\usepackage{multicol}
\usepackage{booktabs}
\usepackage{footmisc}
\usepackage{hhline}
\usepackage{url}
\usepackage{lipsum} 
\usepackage{footmisc}
\usepackage[dvipsnames, table]{xcolor}
\usepackage[colorlinks=true,      
            linkcolor=black,      
            citecolor=black,      
            urlcolor=black]       
            {hyperref}

\IEEEoverridecommandlockouts                              

\newtheorem{ass}{\textbf{Assumption}}

\newtheorem{dnt}{\textbf{Definition}}

\newtheorem{prob}{\textbf{Problem}}

\graphicspath{{figures/}} 
\title{\LARGE \bf
Viability-Preserving Passive Torque Control} 

\author{Zizhe Zhang$^{\dag \ddag 1}$, Yicong Wang$^{\dag 1}$, Zhiquan Zhang$^2$, Tianyu Li$^{1}$, Nadia Figueroa$^{1}$
\thanks{$^{\dag}$Equal Contributions.}
\thanks{$^{\ddag}$Corresponding author. E-mail: \href{mailto:zizhez@seas.upenn.edu}{\textcolor{blue!80!black}{zizhez@seas.upenn.edu}}}
\thanks{$^{1}$University of Pennsylvania, $^{2}$UIUC.} 
}

\begin{document}

\maketitle
\thispagestyle{empty}
\pagestyle{empty}
\begin{abstract}
Conventional passivity-based torque controllers for manipulators are typically unconstrained, which can lead to safety violations under external perturbations. In this paper, we employ viability theory to pre-compute safe sets in the state-space of joint positions and velocities. These viable sets, constructed via data-driven and analytical methods for self-collision avoidance, external object collision avoidance and joint-position and joint-velocity limits, provide constraints on joint accelerations and thus joint torques via the robot dynamics. A quadratic programming-based control framework enforces these constraints on a passive controller tracking a dynamical system, ensuring the robot states remain within the safe set in an infinite time horizon. We validate the proposed approach through simulations and hardware experiments on a 7-DoF Franka Emika manipulator. In comparison to a baseline constrained passive controller, our method operates at higher control-loop rates and yields smoother trajectories.
\textit{Project Website: \href{https://vpp-tc.github.io/webpage/}{\textcolor{magenta}{vpp-tc.github.io}}}
\end{abstract}
\vspace{-2.5pt}
\section{INTRODUCTION}
Physical human-robot interaction (pHRI) is ubiquitous in industrial manufacturing and domestic environments, where effective operation hinges on close coordination between humans and robots. To operate safely under physical contact, robots are required to admit external perturbations while preserving closed-loop stability. This necessitates the design of passive torque controllers. Passivity is characterized by absorbing at least as much energy as is released. This property endows robots with robustness and stability when subjected to external disturbances \cite{vanderSchaft2017}. The classical impedance/admittance framework ensures passivity under appropriate damping \cite{Colgate1988, Hogan1984}, and port-Hamiltonian methods \cite{9655210} achieve it via energy shaping. The energy-tank framework monitors next exchanged energy, accumulates dissipation in a virtual reservoir, and authorized controller adaptations only with a nonnegative energy budget, making it suitable for passivity-preserving variable-impedance control \cite{7801022}. In scenarios where trajectories are specified via a dynamical-systems (DS) representation \cite{dsbook, Figueroa2022}, \cite{kronander2015passive} proposed a passive torque controller for trajectory tracking. 

Nevertheless, such passivity-based controllers generally lack explicit constraint handling, so collision avoidance and joint-limit satisfaction cannot be guaranteed throughout the entire mission execution. To address this problem, \cite{CPIC} (CPIC) leverages the Control Barrier function (CBF) framework \cite{ames2019cbf} and formulates a hierarchically prioritized quadratic program (QP) with hard and soft constraints that constrains a passive torque controller. Within this formulation, joint-limit, self-collision avoidance, external object collision avoidance, and singularity avoidance are treated as safety-critical constraints by defining joint-space analytic or data-driven barrier functions and encoding them as torque-level constraints via Exponential Control Barrier Functions (ECBF) \cite{nguyen2016exponential}. Nonetheless, although this control framework explicitly constrained the torques generated by the passive controller, it has several drawbacks. Although the constraints are partitioned into soft and hard categories, mutual conflicts among the hard constraints can still occur, rendering the QP infeasible. In addition, the QP framework is susceptible to deadlock and entrapment in local minima in the joint space. A further computational challenge arises when determining the parameters of the ECBF: because the safety boundary is learned from data and defined at the joint level while the constraints are imposed at the torque level, one must compute the Hessian of the boundary function with respect to the joint variables. When this boundary is represented by neural networks, as in \cite{CPIC}, obtaining real-time second-order information becomes challenging.

\begin{figure}[!tbp]
    \centering
    \includegraphics[width=1\linewidth]{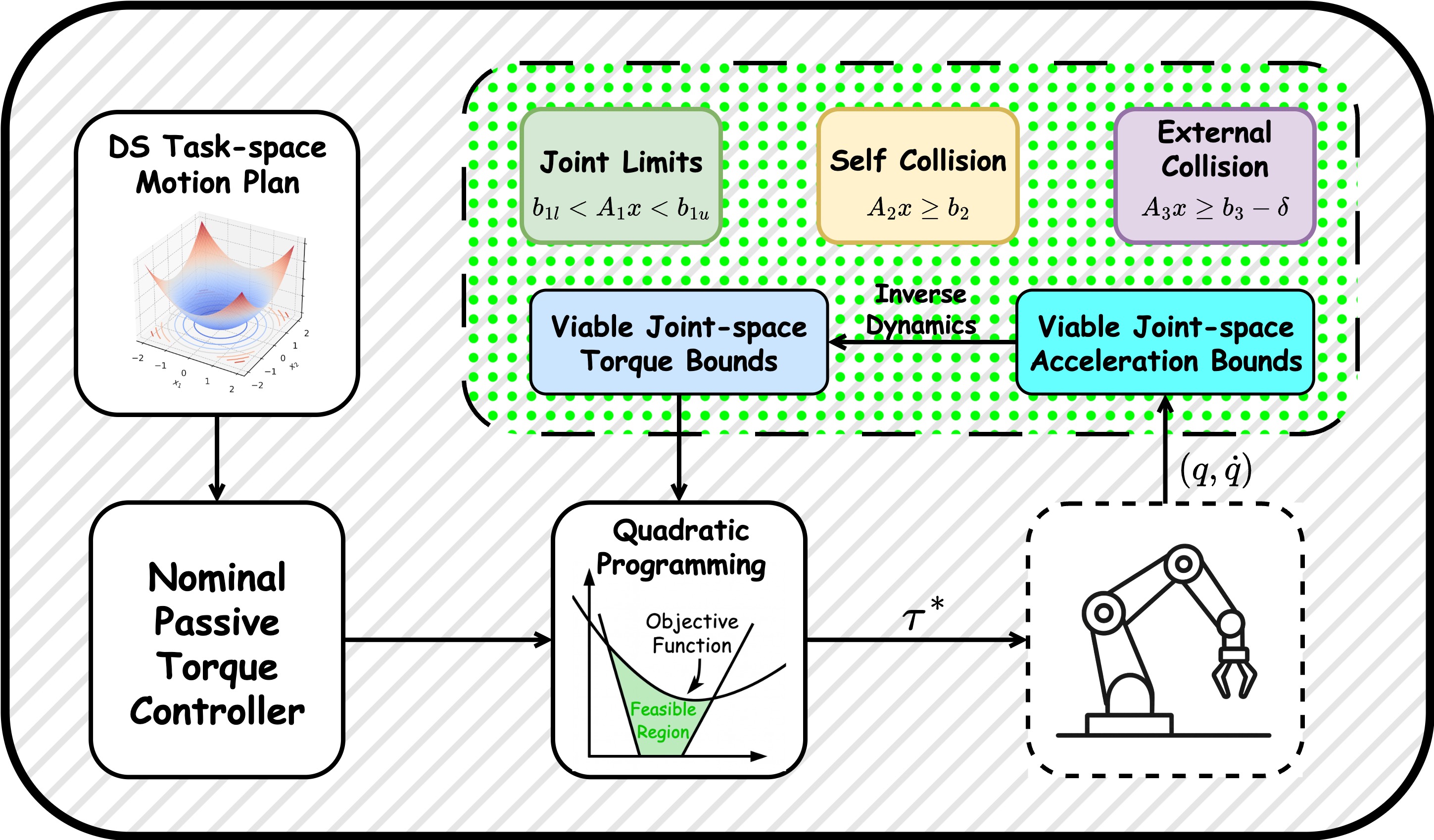}
    \vspace{-15pt}
    \caption{Schematic of the proposed viability-preserving, passivity-based torque controller. Joint-limit torque bounds are derived analytically, whereas bounds for self-collision and external-object collisions are data-driven. \label{fig:control pipeline}}
    \vspace{-15pt}
\end{figure}

\begin{figure*}[!tbp]
    \centering
    \includegraphics[width=1\linewidth]{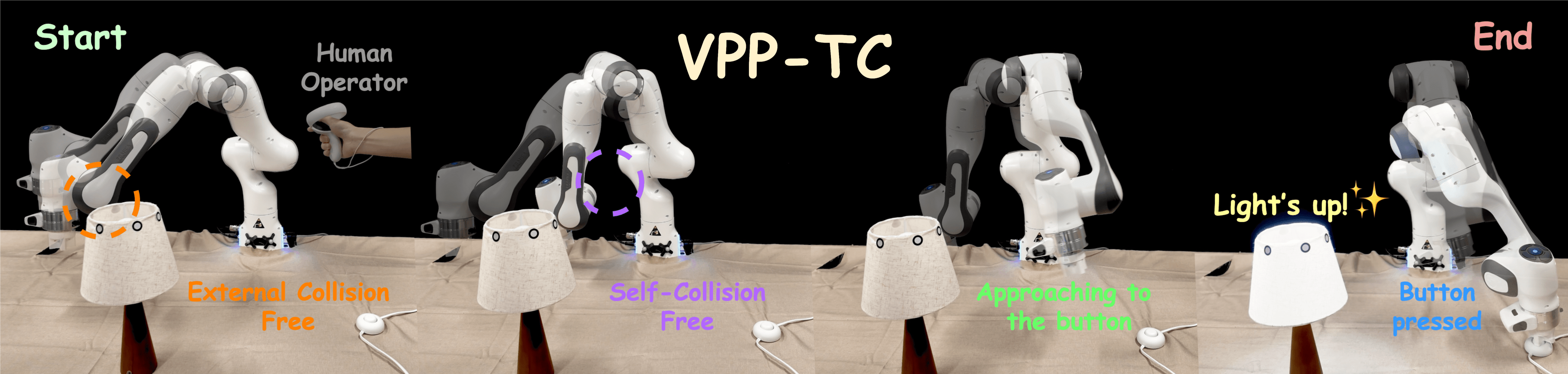}
    \caption{Real-world teleoperation task - pressing a desk lamp button to turn it on. The human operator provides only coarse target positions (no precomputed collision-free trajectory). Our proposed torque controller \textbf{VPP-TC} ensures safety near the lamp and prevents self-collision while accomplishing the task.\label{fig:teleop_lamp}}
    \vspace{-10pt}
\end{figure*}

\textbf{Contributions} In this work, inspired by \cite{bouguerra2019viability,CPIC,viabiliy}, we introduce safety bounds in an augmented state space comprising joint positions and joint velocities. Operating in this space allows us to derive bounds on joint accelerations, and thus on joint torques via the robot dynamics, without requiring second-order derivatives of the boundary functions. We formalize these bounds using the control-theoretic notion of viability, which guarantees controlled forward invariance over an infinite time horizon for some control input sequences. We construct viable set for self-collision and external-object collision avoidance through data-driven methodologies, and we obtain viable sets for joint-position and joint-velocity limits through analytical derivations (in the spirit of \cite{viabiliy}). Building on these sets, we develop a passivity-preserving torque-control framework (Fig.~\ref{fig:control pipeline}) that maintains state viability by solving a QP with torque-level constraints induced by the viable sets; which activate adaptively as the state approaches the corresponding viability boundaries.

\section{PRELIMINARIES}\label{section3}
\subsection{Dynamical System Motion Planning}
We adopt a Dynamical System (DS) formulation for task-space planning \cite{dsbook}. Let $x \in \mathbb{R}^d$ denote the task-space state, whose nominal evolution is governed by the autonomous ODE $\dot{x}=f(x),  f:\mathbb{R}^d\to\mathbb{R}^d$, where $f$ is continuously differentiable and stable attractor $x^* \in \mathbb{R}^d$, i.e.,
\begin{equation}
    \exists \delta_f,\ {\rm s.t.}\ \lim_{t \rightarrow \infty}\|x(t) - x^*\| \leq \delta_f,
\end{equation}
where $\delta_f$ is a positive real number. Accordingly, we set the desired task-space velocity to $\dot x_d = f(x)$.
\subsection{Passivity-Based Control with Dynamical Systems \cite{kronander2015passive} }
\label{section3b}
To enforce task-space passivity while following a desired Dynamical System (DS) motion plan, we use a velocity-based impedance controller:
\begin{equation}
    F_c = G_x(x) - D(x)(\dot x - f(x)),
\end{equation}
where $G_x(x) = J(q)^{-\top}G(q)$ is the gravity term expressed in task space, $D(x) \in \mathbb{R}^{d\times d}$ is a state-dependent damping matrix, $\dot x$ is measured task-space velocity, and $f(x)$ is the desired DS velocity. Intuitively, the gravity term compensates potential energy due to gravity, and the damping term applies anisotropic dissipation to the velocity error. With a suitable $D(x)$, the controller injects power only along the desired motion $f(x)$ and dissipates power in directions orthogonal to it, which yields passive behavior while tracking the DS.
\begin{ass}
    The vector field $f(x)$ is the gradient flow of a scalar potential $\mathcal{P}(x)$, that is, $f(x) = -\nabla_x \mathcal{P}(x)$.
\end{ass}
The damping matrix is positive semidefinite and factorizes as $D(x) = V(x)\Lambda V(x)^\top$, with $\Lambda$ a diagonal matrix of nonnegative entries. The columns of $V(x)$ form an orthonormal frame chosen so that $v_1 = f(x)/\|f(x)\|$ aligns with the desired direction of motion, while the remaining columns span its orthogonal complement.

Under these conditions, the negative velocity-error feedback steers $\dot x$ towards $f(x)$, producing kinetic energy only along the DS direction and dissipating it elsewhere, thereby achieving task-space passivity during tracking.

\subsection{Safety Guaranteed by Viability}
\begin{figure}[h]
    \vspace{-10pt}
    \centering
    \includegraphics[width=0.7\linewidth]{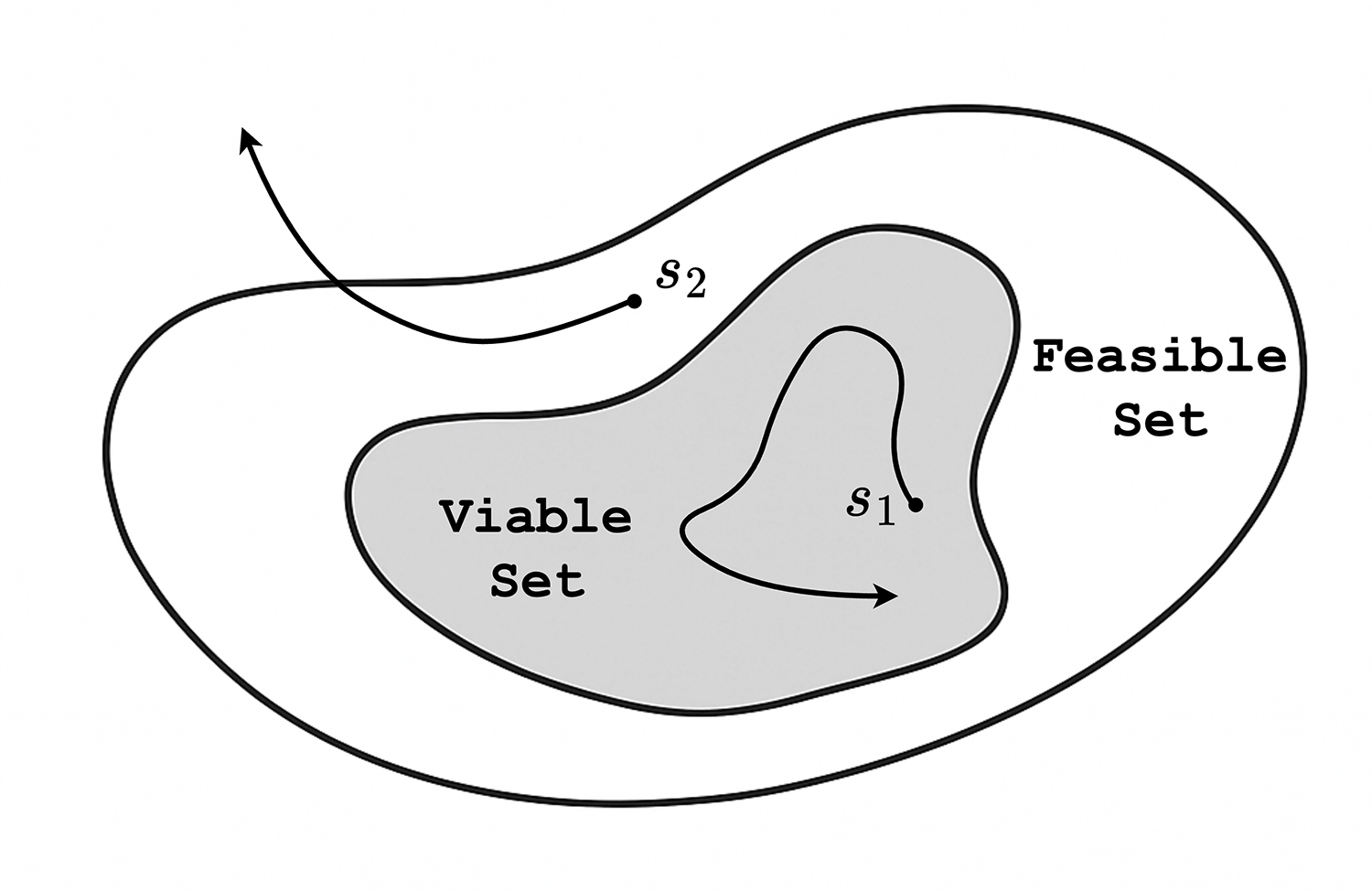}
    \vspace{-10pt}
    \caption{Viability concept:  
    $s_1$ is feasible and viable, whereas $s_2$ is feasible but non-viable.\label{fig:via-concept}}
    \vspace{-5pt}
\end{figure}

We define the set of viable states for an $n$-DoF manipulator as the Cartesian product of a set of $q\in \mathbb{R}^n$ (joint positions) and $\dot q\in \mathbb{R}^n$ (joint velocities), $(q, \dot q) \in \mathcal{V}$. A state $(q, \dot q) \in \mathbb{R}^{2n}$ is viable if from any initial state within $\mathcal{V}$, there exists a control sequence $\ddot q$ (acceleration) that generates an infinite state sequence which is inside the viable set, whereas a merely feasible state may leave the feasible set under future control sequences (Fig.~\ref{fig:via-concept})\cite{bouguerra2019viability,viabiliy}. We formally define the viable states and viable set as:
\begin{dnt}[Viability]
    \begin{equation}\label{eq:via}
\begin{aligned}
 (q(0), \dot q(0)) \in \mathcal{V} 
 \;\Leftrightarrow\; 
   \exists (\ddot q_i)_{i=0}^\infty : (q(t), \dot q(t))\in \mathcal{F}, \;\forall t \ge 0&; \\
 |\ddot{q}_k| \leq \ddot{q}_k^{\max}, \; k = 1,2, \ldots , n&,
\end{aligned}
\end{equation}
where $\ddot q_k^{\max}$ represents hardware joint acceleration limits of the $k$-th joint. $\mathcal{F} \in \mathbb{R}^{2n}$ denotes the feasible set of state pairs $(q, \dot q)$ that satisfy all constraints.
\end{dnt}
We define the safe set as the viable state set $\mathcal{V}$.
\begin{ass}
    The initial state $(q_0, \dot q_0)$ is assumed to be within the viable state set $\mathcal{V}$.
\end{ass}
\section{PROBLEM FORMULATION}\label{section2}
We seek to control an $n$-DoF manipulator with rigid-body dynamics derived by the Euler-Lagrange equation:
\begin{equation}\label{eq:dyna}
    M(q)\ddot q + C(q, \dot q) \dot q + G(q) = \tau_c + \tau_{ext},
\end{equation}
where $\ddot q \in \mathbb{R}^n$ denote joint accelerations. $M(q) \in \mathbb{R}^{n \times n}$, $C(q, \dot q) \in \mathbb{R}^{n \times n}$, $G(q) \in \mathbb{R}^{n}$ denote the inertia matrix, Coriolis matrix, and gravity respectively. $\tau_c \in \mathbb{R}^n$ and $\tau_{ext} \in \mathbb{R}^n$ denote the control input and external torque applied to the joints.

The goal of this paper is to provide torque bounds on nominal passive torque controllers \cite{kronander2015passive}. The torque bounds are intended to guarantee viability \cite{bouguerra2019viability,viabiliy} for joint space constraints, whose feasible sets are formulated as
\begin{itemize}
    \item Joint Position and Velocity Limits:
    \begin{equation*}
        \mathcal{F}_{\text{jnt}} := \left\{ (q, \dot q)\,|\, q^- \leq q \leq q^+, \left|\dot{q}\right| \leq \dot{q}_{\max} \right\}.
    \end{equation*}
    where $q^-, q^+ \in \mathbb{R}^n$ denote the per-joint lower/upper position bounds, 
$\dot q_{\max}\in \mathbb{R}^n_{+}$ denotes the per-joint maximum velocity, 
and the inequalities are interpreted elementwise.
    \item Self-Collision Avoidance: 
    \begin{equation*}
        \mathcal{F}_{\text{sca}} := \{(q, \dot q)\,|\,q \in \mathcal{C}_{\text{sca}} \},
    \end{equation*}
    where $\mathcal{C}_{\text{sca}}\subset\mathbb{R}^n$ is the set of self-collision-free configurations.
    \item External Obstacle Collision Avoidance: 
    \begin{equation*}
        \mathcal{F}_{\text{eca}} := \{(q, \dot q)\,|\,q \in \mathcal{C}_{\text{eca}} \},
    \end{equation*}
    where $\mathcal{C}_{\text{eca}}\subset\mathbb{R}^n$ is the set of external-collision-free configurations.
\end{itemize}
According to \eqref{eq:via}, we can then obtain the separated viable state set  $\mathcal{V}_{\text{jnt}}$, $\mathcal{V}_{\text{sca}}$ and  $\mathcal{V}_{\text{eca}}$. The final viable state set is given by the intersection $\mathcal{V} = \mathcal{V}_{\text{jnt}}\,\cap\,\mathcal{V}_{\text{sca}}\,\cap\,\mathcal{V}_{\text{eca}}$ which is defined as the safe state set.

\begin{prob}
    Consider a torque controlled manipulator with dynamics \eqref{eq:dyna}. Design a passive torque controller that guarantees safety by preserving the state $(q, \dot q)$ within the viable state set $\mathcal{V}$.
\end{prob}

\begin{figure}[t]
  \centering
  \includegraphics[width=\linewidth]{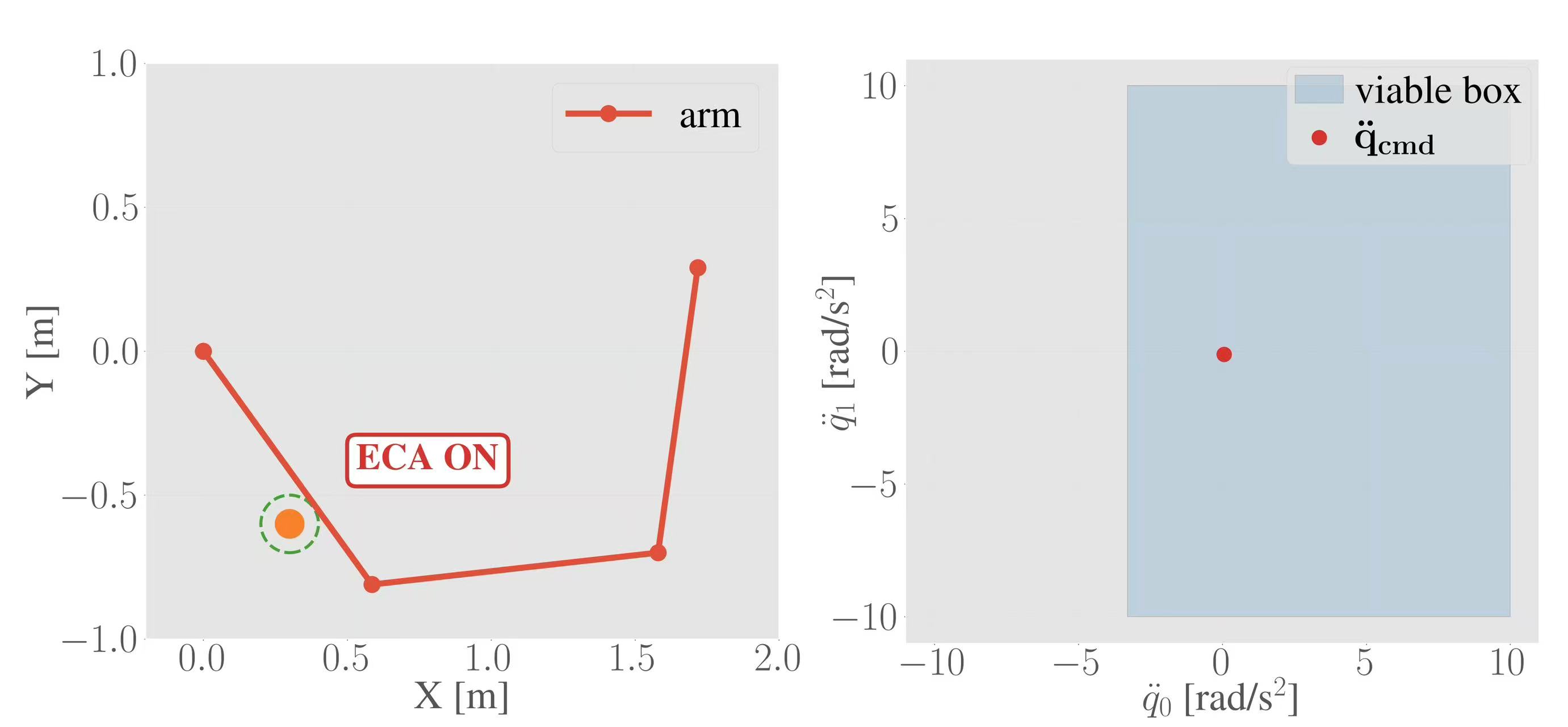}
  \caption{Planar 3-DoF demo. Left: workspace trajectory. Right: viable box of \((\ddot q_0,\ddot q_1)\) under hardware limits \([-10,10]\ \mathrm{rad/s^2}\). With ECA enabled, the admissible box contracts as the obstacle constraint activates, illustrating viability-induced acceleration bounds.\label{fig:rtb_viable_boxes}}
  \vspace{-10pt}
\end{figure}



\section{PROPOSED APPROACH}
Our proposed approach handles the three aforementioned forms of collision avoidance in a unified manner by testing whether the current state belongs to the viable set. Whenever the state is outside the viable set, we formulate and solve an optimization problem to return it to the viable set. We refer to this method as the \textit{Viability-Preserving Passive Torque Controller (VPP-TC)}, which we will use for brevity throughout the remainder of the paper.
A planar 3-DoF example illustrating viability-induced acceleration bounds is shown in Fig.~\ref{fig:rtb_viable_boxes}. The viable box for $(\ddot q_0,\ddot q_1)$ shrinks from the hardware limits $[-10,10]\ \mathrm{rad/s^2}$ as the ECA constraint activates.
\subsection{Self-Collision Avoidance Constraints}\label{sca-cons}
We conservatively define a joint state as self-collision-free viable if, under maximum deceleration opposite to the current joint velocity, the robot can come to a complete stop without any self-collision throughout the braking trajectory.
\begin{dnt}[Self-collision-free viability]
Let $\mathcal{C}_{\text{sca}}\subset\mathbb{R}^n$ be the set of self-collision-free configurations.
Given an initial state $(q_0,\dot q_0)\in\mathbb{R}^{2n}$, define the \emph{braking control}
$u^{\text{br}}(t)$  
that applies maximum deceleration opposite to the current velocity, element-wise, until stop,
and let $(q(t),\dot q(t))$ be the corresponding trajectory.
The \emph{braking time} $T_{\text{br}}(q_0, \dot{q}_0)$ is the first time such that $\|\dot q(T_{\text{br}})\|=0$,
and the \emph{braking trajectory set} is
\begin{equation}\label{traj:br}
    \mathcal{C}_{\text{br}}(q_0, \dot{q}_0) : =  \{\, (q(t),\dot q(t))\ ,\forall t\in[0,\,T_{\text{br}}] \,\}.
\end{equation}
We then define the conservative self-collision viable set as
\[
\mathcal{V}_{\text{sca}}
:=
\big\{\, (q_0,\dot q_0)\ \big|\ 
q(t)\in \mathcal{C}_{\text{sca}},\ \forall\, t\in[0,\,T_{\text{br}}]\big\}.
\]

\end{dnt}

\textbf{Learning $\mathcal{V}_{\text{sca}}$:} To learn such conservative viability from data, we employ a transformer-based neural network to map states $(q,\dot q)$ to a real value $\Gamma(q,\dot q) : \mathbb{R}^{2n} \rightarrow \mathbb{R}$. The network first applies a linear projection and learned positional encodings to the input $(q,\dot q)$, followed by a multi-layer transformer encoder that processes the sequence of state embeddings. The output is aggregated via global average pooling and passed through an MLP classifier with GeLU activation to produce the logits $(\ell_1, \ell_2)$, from which $\Gamma(q,\dot q)$ is defined as $\Gamma(q,\dot q) = \ell_1 - \ell_2$.

This mapping is used to to define an \emph{empirical surrogate} of the viable state set
\[
\hat{\mathcal{V}}_{\text{sca}}
:= \{(q,\dot q)\ |\ \Gamma(q,\dot q) > 0\}.
\]
For scenarios where $\Gamma(q,\dot q) \in (0, \epsilon_{\text{sca}}]$, 
with $\epsilon_{\text{sca}} \in \mathbb{R}_{++}$ denoting a small positive scalar, we impose a constraint enforcing $\Delta \Gamma(q,\dot q) \ge 0$, thereby ensuring that the state remains within the viable state set.
By Taylor expansion and omitting high-order terms, we have
\begin{equation}
\begin{aligned}
    \Delta \Gamma(q, \dot q) =& \nabla_q \Gamma^\top \cdot \Delta q + \nabla_{\dot q}\Gamma^\top \cdot \Delta \dot q\\
    =& \nabla_q \Gamma^\top \cdot (\dot q \delta t + \frac{1}{2}\ddot q\delta t^2) + \nabla_{\dot q}\Gamma^\top \cdot (\ddot q \delta t)\\
    =& (\frac{1}{2}\nabla_q \Gamma^\top \cdot \delta t^2 + \nabla_{\dot q} \Gamma^\top \cdot \delta t)\ddot q + \nabla_q \Gamma^\top \cdot \dot q \delta t,
\end{aligned}
\end{equation}
which is an affine function of $\ddot q$. For simplicity, we define $g_{se} = \frac{1}{2}\nabla_q \Gamma^\top  \cdot \delta t^2 + \nabla_{\dot q}\Gamma^\top  \cdot \delta t$ and $b_{s} = \nabla_q \Gamma^\top  \cdot (\dot q \delta t)$. Then we have the constraint of $\ddot q $ as $g_{se} \ddot q + b_{s} \ge 0$. Plugging this acceleration constraint into the robot dynamics \eqref{eq:dyna}, we have the torque constraint as:
\begin{equation}
    g_{se} M^{-1} \tau \ge g_{se} M^{-1}(C\dot q+G) - b_{s}
\end{equation}

\subsection{External Obstacle Collision Avoidance Constraints}
We employ the Bernstein Polynomial representation~\cite{li2023learning} to encode the SDF of every robot link.  
Let $\Omega_n$ denote the $n$‑th link of a robot, for a query point $ p\in\mathbb R^{3}$ expressed in the base frame, the distance from $p$ to the entire robot surface at configuration $q$ is the pointwise minimum over all links:
\begin{equation}
  S(p,q)=
  \min_{n=1,\dots,N} S_{\Omega_n}^{b}(p,q).
  \label{eq:sdf_robot}
\end{equation}

Then we define the viability-preserving SDF 
as the minimum
instantaneous SDF encountered along the braking trajectory $\mathcal{C}_{\text{br}}(q_0,\dot q_0)$ defined in subsection~\ref{sca-cons}:
\begin{equation}
\label{eq:sv_cont}
S_v(p, q_0,\dot q_0)
\;=\; \min_{t\in[0,T_{\text{br}}]} S\big(p, q(t)\big).
\end{equation}
Following \cite{li2023learning}, we replace the nondifferentiable $\min$ in Eqs.~\ref{eq:sdf_robot}-\ref{eq:sv_cont} with a differentiable smooth approximation; in our implementation we use a soft-min (log-sum-exp) surrogate to ensure differentiability and avoid gradient discontinuities.

\begin{dnt}[External-collision-free viability]
    Similar to subsection~\ref{sca-cons}, we define the conservative external-collision viable set as
    \[
        \mathcal{V}_{\text{eca}}
        :=
        \big\{\, (q_0,\dot q_0)\ \big|\ 
        q(t)\in \mathcal{C}_{\text{eca}},\ \forall\, t\in[0,\,T_{\text{br}}]\big\}.
    \]  
\end{dnt}
Also, we use the viability-preserving SDF $S_v(p, q,\dot q)$ to define an \emph{empirical surrogate} of the viable state set
\[
\hat{\mathcal{V}}_{\text{eca}}
:= \{(q,\dot q)\ |\ S_v(p, q,\dot q) > 0\},
\]

For scenarios where $S_v(p, q,\dot q) \in (0, \epsilon_{\text{eca}}]$, with $\epsilon_{\text{eca}} \in \mathbb{R}_{++}$ denoting a small positive scalar, we impose a constraint enforcing $\Delta S_v(p, q,\dot q) \ge 0$, thereby ensuring that the state remains within the viable state set.
By taking the Taylor expansion of $S_v(p, q, \dot q)$ and omitting high-order terms, we have
\begin{equation}\label{eq:delsc}
    \begin{aligned}
        \Delta S_v(p, q,\dot q)
        =& \nabla_q S_v^\top \cdot \Delta q + \nabla_{\dot q}S_v^\top \cdot \Delta \dot q + \nabla_p S_v^\top \cdot\Delta p\\
        =& \nabla_q S_v^\top\cdot(\dot q\delta t + \frac{1}{2}\ddot q \delta t^2) + \nabla_{\dot q}S_v^\top\cdot(\ddot q \delta t) \\
        &+ \nabla_p S_v^\top \cdot\Delta p\\
        =& (\frac{1}{2}\nabla_q S_v^\top\cdot \delta t^2 + \nabla_{\dot q}S_v^\top\cdot \delta t)\ddot q\\& + \nabla_q S_v^\top\cdot(\dot q \delta t) + \nabla_p S_v^\top\cdot \Delta p.
    \end{aligned}
\end{equation}
For simplicity, we define $g_{ee} = \frac{1}{2}\nabla_q f^\top  \cdot\delta t^2 + \nabla_{\dot q} f^\top \cdot\delta t$ and $b_e = \nabla_q f^\top \cdot(\dot q\delta t) + \nabla_p f^\top\cdot \Delta p$. The constraint is then expressed as $g_{ee}^\top \ddot q + b_e \ge 0$, which is also affine to $\ddot q$. Plugging this acceleration constraint into dynamics \eqref{eq:dyna}, we have the torque constraint as
\begin{equation}
    g_{ee} M^{-1} \tau \ge g_{ee} M^{-1} (C \dot q + G) - b_e
\end{equation}

\begin{figure}[!th]
    \centering
\includegraphics[width=0.90\linewidth]{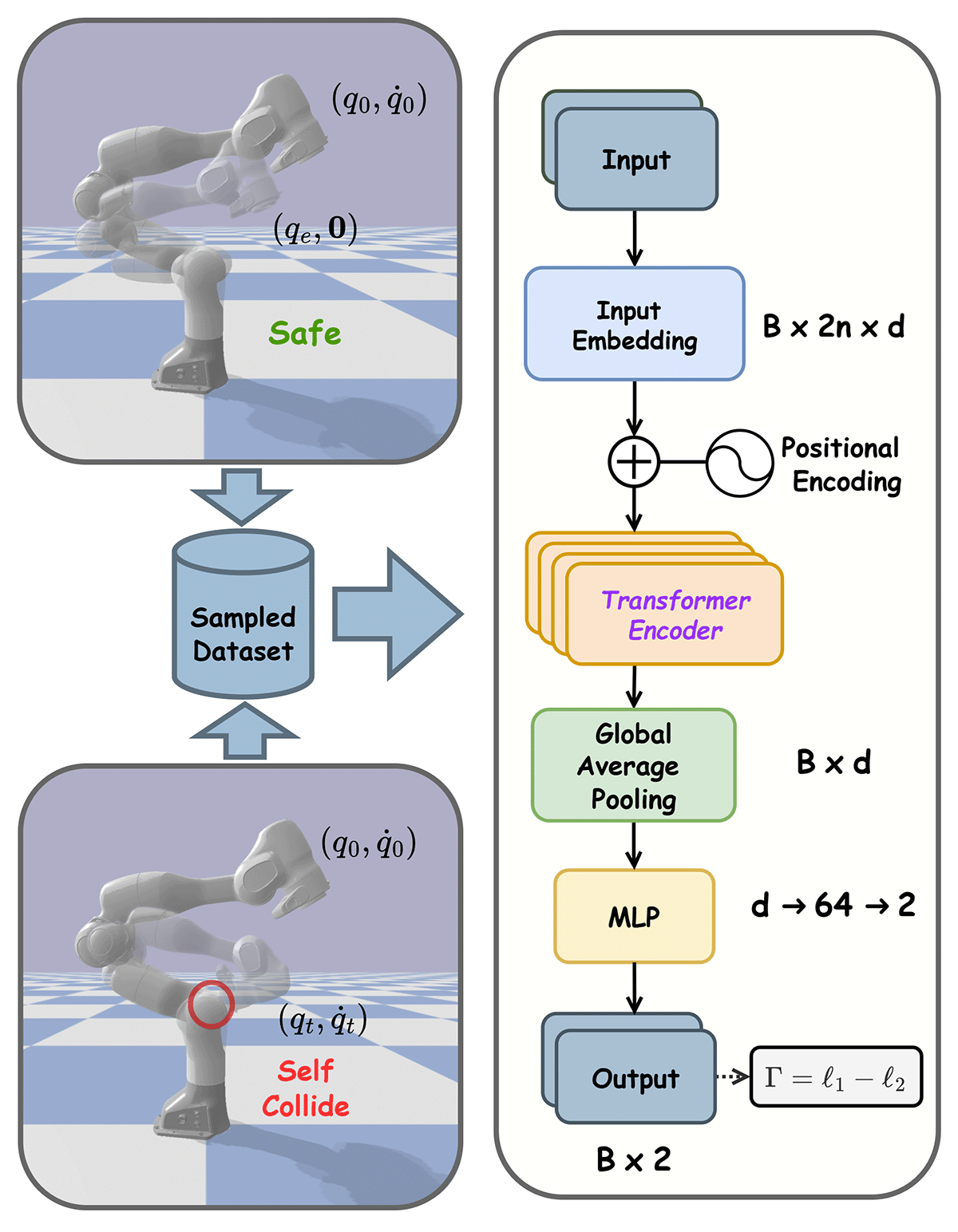}
\vspace{-5pt}
    \caption{Braking rollout from $(q_0,\dot q_0)$ to rest $(q_e,\mathbf{0})$: label \textbf{Safe} iff every state along the trajectory to $(q_e,\mathbf{0})$ is self-collision-free; otherwise \textbf{Self-Collide} at the first contact $(q_t,\dot q_t)$. The resulting dataset trains a network that maps $(q,\dot q)$ to a viability score $\Gamma$.
    \label{fig:sc_model}}
    \vspace{-15pt}
\end{figure}

\subsection{Joint Position and Velocity Limit Constraints}

In this subsection, inspired by \cite{viabiliy}, we describe the formulation of joint position and velocity limits constraints using the definition of viability. We aim to find the bounds of torques to preserve the state $(q, \dot q)$ within $\mathcal{V}_{\rm jnt}$. 
The joint acceleration bound is obtained by Algorithm 1 \cite{viabiliy} for position inequalities, 
\begin{equation}
    \frac{1}{\delta t} (-\dot q_{\max} - \dot q) \leq \ddot q \leq \frac{1}{\delta t}(\dot q_{\max}-\dot q)
\end{equation}
for velocity inequalities, and
the intersection of 
\begin{equation}
    \begin{aligned}
        \dot q + \delta t \ddot q &\leq \sqrt{2\ddot q_{\max}(q^+-q-\delta t\dot q-0.5\delta t^2\ddot q)}\\
        \dot q + \delta t \ddot q &\ge -\sqrt{2 \ddot q_{\max}(q + \dot q \delta t + 0.5 \ddot q \delta t^2)-q^-}
    \end{aligned}
\end{equation}
for viability inequalities.
By taking the intersection of these three sets of inequalities with the hardware acceleration limits $\ddot q_{max}$, we obtain a maximal lower bound $\ddot q_{lb}$ and a minimal upper bound $\ddot q_{lb}$ such that
\begin{equation}\label{eq:jb}
    \ddot q_{lb} \leq \ddot q \leq \ddot q_{ub}
\end{equation}
For further details, please refer to \cite{viabiliy}. 

We can then obtain the torque constraints by plugging \eqref{eq:jb} into the dynamics, which yields
\begin{equation}
    \ddot q_{lb} + M^{-1}(C \dot q + G)\leq M^{-1} \tau \leq \ddot q_{ub} + M^{-1}(C \dot q + G)
\end{equation}
\subsection{Passive Torque Control with All Constraints}
In our control framework, the joint position and velocity limit constraints are always active, the self-collision avoidance constraints and external obstacle collision avoidance constraints are active when the state is close to the boundary of the viable set. The optimization problem is formulated as:
\begin{equation}\small\label{eq:QP_framework}
    \begin{aligned}
    \min_{\tau} \quad & \frac{1}{2}\|\hat J(q)^{-\top}\tau - F_c(x)\|_2^2 + \alpha_1 \|\tau\|_2^2 + \alpha_2\|\delta\|_2^2\\
    \textrm{s.t.} \quad & \ddot q_{lb} + M^{-1}(C \dot q + G)\leq M^{-1} \tau \leq \ddot q_{ub} + M^{-1}(C \dot q + G)\\
    \quad & g_{se} M^{-1} \tau \ge g_{se} M^{-1}(C\dot q+G) - b_{s}\ \wedge (\Gamma \in (0, \epsilon_{\text{sca}}])\\
    \quad & g_{ee} M^{-1} \tau \ge g_{ee} M^{-1} (C \dot q + G) - b_e-\delta\ \wedge (S_v \in (0, \epsilon_{\text{eca}}])\\
    &\tau \in [\tau_{\min}, \tau_{\max}]    \\
    & \delta \ge 0
\end{aligned}
\end{equation}
We define $\hat J(q)^\top$ as $((J(q)J(q)^\top)^{-1}+ \sigma^2 \mathbb{I}_n)J(q)$ to avoid singular configurations, $\sigma$ denotes a small real number and $\mathbb{I}_n$ denotes an identity matrix. $\delta$ is a non-negative slack variable.
Passivity analysis see in Appendix~\ref{appdx-a}.

\begin{figure*}[!tbp]
    \centering
    \subfigure[With the SCA, the learned score $\Gamma(q,\dot q)$ and the $\min$ link–link distance remain above the threshold; without SCA the robot enters self-collision.\label{fig:sim-sca}]{
        \includegraphics[width=0.315\linewidth]{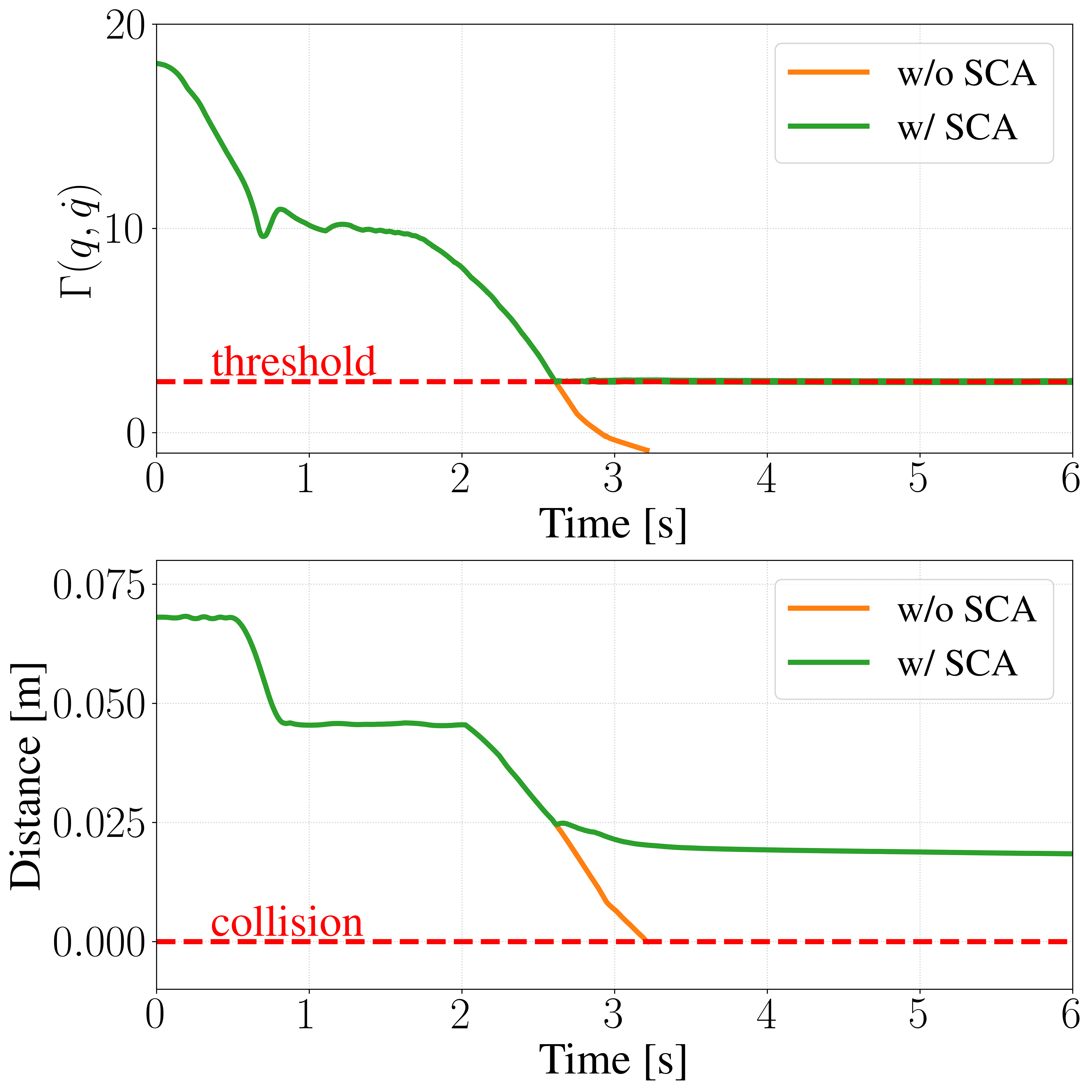}
    }\hfill
    \subfigure[With the ECA, the viability-preserving distance $S_v(p,q,\dot q)$ and the clearance to the obstacle stay positive; without ECA they collide.\label{fig:sim-eca}]{
        \includegraphics[width=0.315\linewidth]{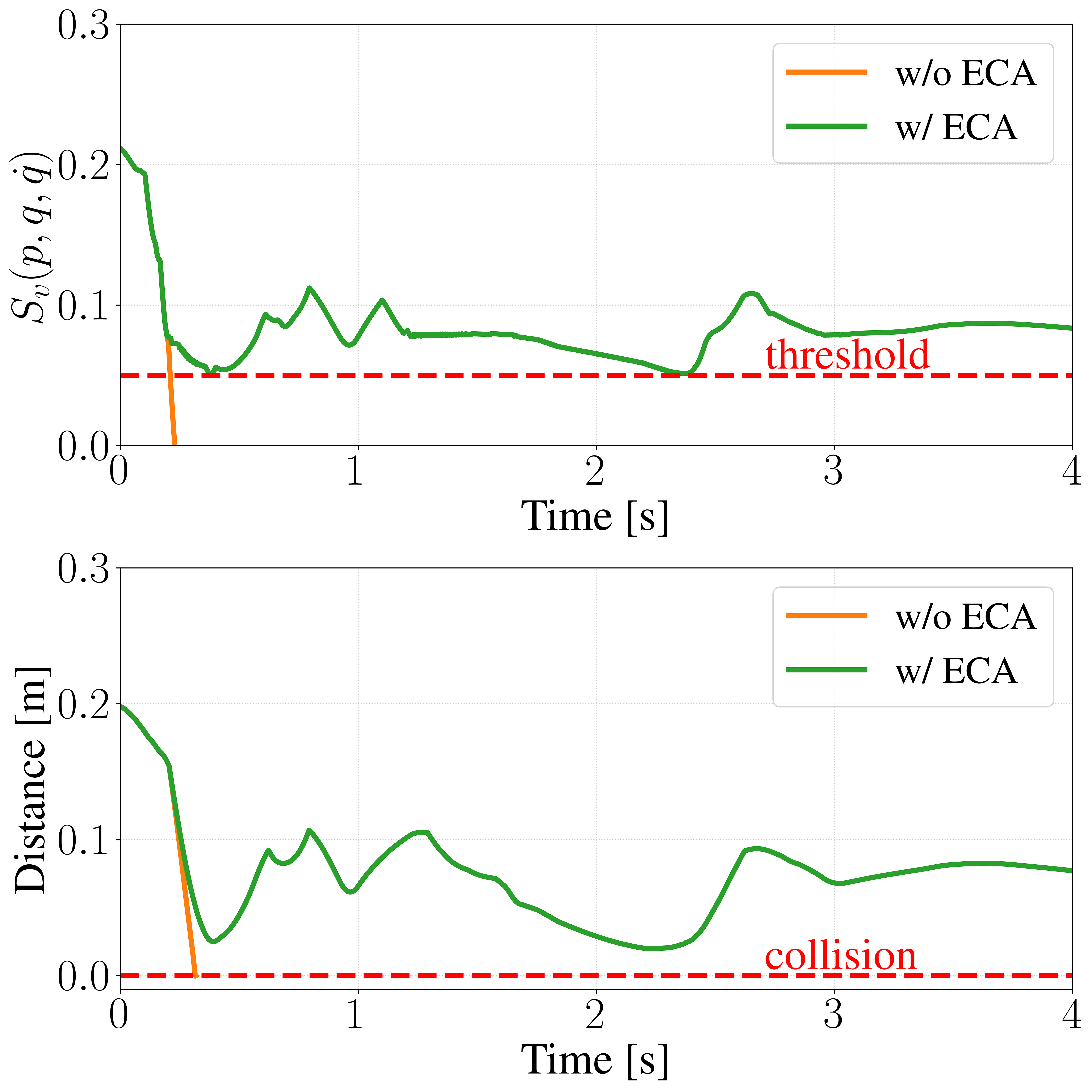}
    }\hfill
    \subfigure[With both SCA and ECA active, the controller maintains positive distances to self-collision and external obstacles throughout the trajectory.\label{fig:sim-all}]{
        \includegraphics[width=0.315\linewidth]{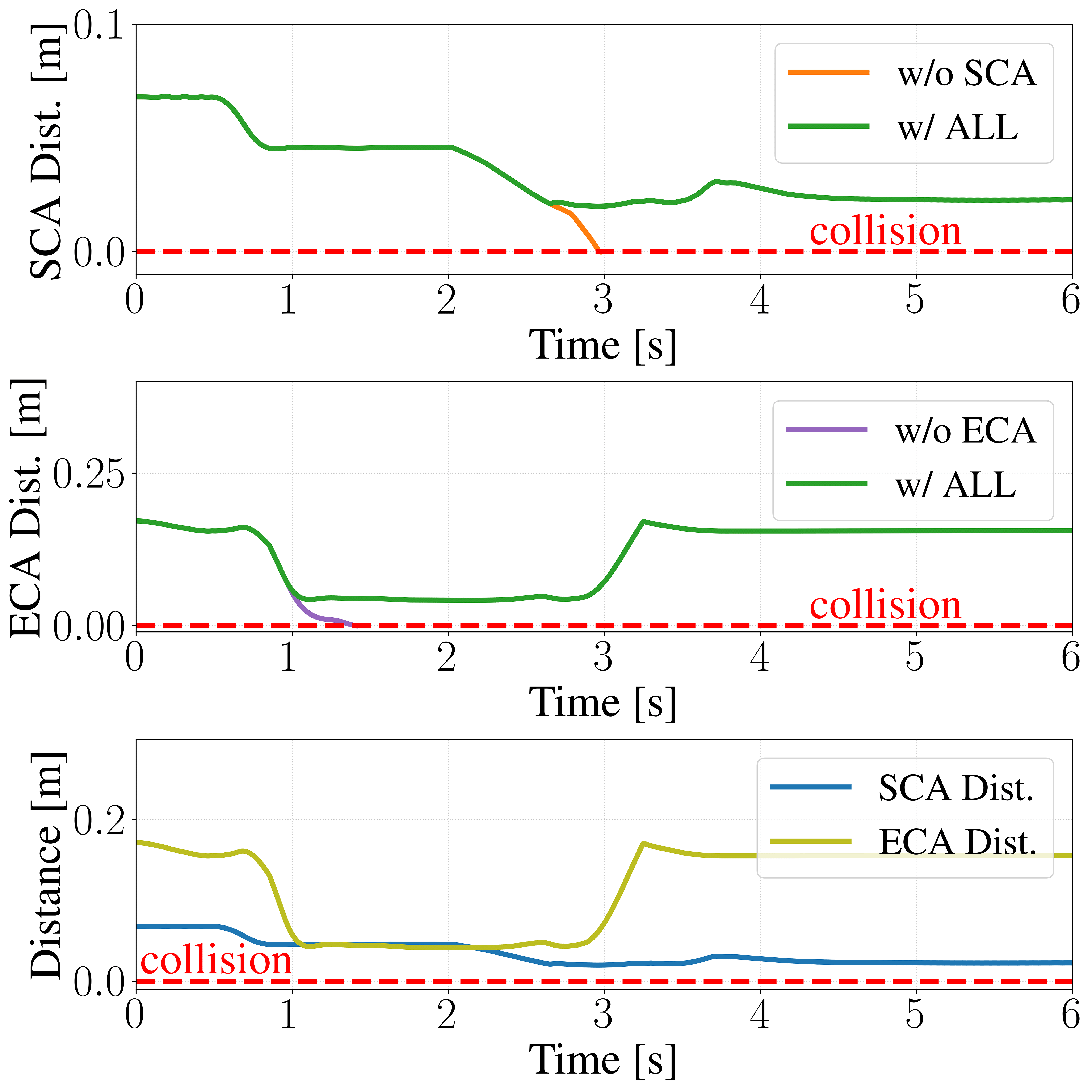}
    }
    \caption{Simulation results of VPP-TC under different safety constraints.\label{fig:sim}}
    \vspace{-10pt}
\end{figure*}

\section{EXPERIMENTAL RESULTS}
\label{section5}
All experiments were conducted using a 7-DoF Franka Panda manipulator. 
For simulations, we employed PyBullet and solved the QP optimization problem using CVXPY~\cite{diamond2016cvxpy}, 
while CVXGEN~\cite{MattingleySB12} was utilized for real-robot experiments. 
Videos of both PyBullet simulations and hardware experiments are included in the supplementary material
and available on our project \href{https://vpp-tc.github.io/webpage/}{website}.

\subsection{Network Training}
We train two learning models used by the controller:
(i) a transformer-based classifier $\Gamma(q,\dot q)$ that predicts conservative self-collision viability, described in Sec.~\ref{sca-cons}, and
(ii) per-link SDFs represented with Bernstein polynomials \cite{li2023learning} that enable fast and differentiable distance queries for external-collision reasoning, used in the computation of the viability-preserving distance $S_v$ in \eqref{eq:sv_cont}.

\textit{1) Self-collision network:}
We generate 3 million labeled pairs $(q,\dot q)$ in simulation by rolling out the braking control until the system comes to a full stop. A sample is marked viable if the entire rollout is self-collision-free and non-viable otherwise. This dataset is used to train the transformer-based classifier with a standard Transformer-Encoder backbone~\cite{DBLP:journals/corr/VaswaniSPUJGKP17}: with feed-forward size 128, 2 attention heads per layer, and 4 stacked transformer encoder layers, trained for 30 epochs. On the validation set, we select $\gamma_{\text{thr}}$ via a threshold sweep over $\Gamma$, choosing the value that leads to better recall for the viable class. On the test set at this fixed threshold, the classifier attains 99.27\% accuracy and 99.74\% recall.

\textit{2) External-collision network:}
For each robot link $\Omega_n$, we represent the geometry by an offline Bernstein-polynomial SDF $S_{\Omega_n}^b$ learned from its mesh model~\cite{li2023learning}.
We define a cubic volume around the link; points inside the volume are normalized into $[0,1]$, while points outside are projected onto the boundary. The SDF of a link is parameterized by a trivariate Bernstein basis with $24$ basis functions per axis (total $24^3$), and the coefficient vector is optimized by recursive least squares with ridge regularization, updating the weights incrementally on mini-batches. During inference, all link-level SDFs are transformed to the base frame using the kinematic chain, and the whole-body SDF $S(p,q)$ is computed as in Eq.~\eqref{eq:sdf_robot}. We evaluate external-collision distances along the entire braking trajectory. 
At each discretized step $t\in[0,T_{\mathrm{br}}]$, we query the whole-body SDF $S(p,q(t))$ to measure clearance and take the $\min$ across all time steps. 
The resulting \emph{viability-preserving distance} is defined as Eq.~\eqref{eq:sv_cont}.

\subsection{Simulation Experiments}
\label{section5b}
We designed three simulation experiments on the 7-DoF Franka Panda in PyBullet to evaluate the effectiveness of the proposed viability-preserving passive torque controller under different safety constraints: 1) joint limits with self-collision avoidance, 2) joint limits with external collision avoidance, and 3) joint limits with both self-collision and external collision avoidance. 
The robot is initialized at joint position 
\([\,0.669,\,-0.346,\,-0.742,\,-1.66,\,-0.367,\,2.3,\,1.99\,]^\top\) 
with all joint velocities set to zero. Additionally, we compared our method against the baseline \textit{Constrained Passive Interaction Control (CPIC)}~\cite{CPIC}, highlighting improvements in computational efficiency, path length, and trajectory smoothness. All simulation experiments were performed on a workstation running Ubuntu 20.04, equipped with an AMD Ryzen 7 9800X3D CPU and an NVIDIA GeForce RTX 4060 GPU.

\begin{table*}[h!]
\centering
\caption{Comparison of CPIC and VPP-TC across different constraint settings (average over 5 trials).} 
\renewcommand{\arraystretch}{1.15}
\setlength{\tabcolsep}{5pt}
\begin{tabular}{||c|c|c|c||c|c|c||c|c|c|}
\hline
 & \multicolumn{3}{c||}{Control Loop Rate [Hz]~$\uparrow$} & \multicolumn{3}{c||}{Trajectory Length [$m$]~$\downarrow$}& \multicolumn{3}{c|}{Trajectory Jerkness~$\downarrow$} \\
\cline{2-10}
 & SCA & ECA & ALL & SCA & ECA & ALL & SCA & ECA & ALL \\
\hline\hline
CPIC 
&\(126.1\!\pm\!1.4\)
&\(72.6\!\pm\!0.3\)
&\(65.8\!\pm\!0.2\)
&\(1.12\!\pm\!0.29\)
&\(1.54\!\pm\!0.23\)
&\(1.59\!\pm\!0.23\)
&\(8.8\!\pm\!2.3\)
&\(1913.7\!\pm\!726.0\)
&\(2173.9\!\pm\!788.0\)\\
\hline
\textbf{VPP-TC} 
&\cellcolor{blue!15}$\textbf{171.6}\!\pm\!9.1$&\cellcolor{blue!15}$\textbf{153.0}\!\pm\!4.9$&\cellcolor{blue!15}$\textbf{155.4}\!\pm\!10.3$
&\cellcolor{blue!15} $\textbf{0.75}\!\pm\!0.18$
&\cellcolor{blue!15}$\textbf{1.23}\!\pm\!0.19$
&\cellcolor{blue!15}$\textbf{1.19}\!\pm\!0.21$
&\cellcolor{blue!15}$ \textbf{1.2}\!\pm\!0.4$
&\cellcolor{blue!15} $\textbf{216.2}\!\pm\!262.8$
&\cellcolor{blue!15}$\textbf{186.7}\!\pm\!116.8$\\
\hline
\end{tabular}
\label{tab:comp_cpic}
\vspace{-10pt}
\end{table*}

\textit{1) Joint Position and Velocity Limits \& Self-Collision Avoidance:} In this scenario, the desired task-space DS $f(x)$ has a convergence point located within the manipulator’s own body. Without enforcing the self-collision constraint, the robot would inevitably collide with itself.\\
Thus, we simply define the potential function as
\begin{equation}
  V(x) = (x - x^\ast)^\top P (x - x^\ast),\label{eq:V}
\end{equation}
where \(x^\ast = [\,0,\,0,\,0.3\,]^\top\) and \(P = -25\,\mathbb{I}_n\).
The associated vector field is then given by
\begin{equation}
  \dot x = f(x) = \nabla_x V(x) = 2P\,(x - x^\ast).\label{eq:f}
\end{equation}
In this test, we set a threshold $\gamma_{\text{thr}}=2.5$, which the manipulator is required to remain above. This threshold was selected based on the recall performance of the transformer-based neural network classifier, ensuring a high likelihood of correctly identifying viable states. Fig.~\ref{fig:sim-sca} shows that, without the SCA constraint, the manipulator enters self-collision configurations where link1 and link6 intersect. With the self-collision constraint enforced, all non-adjacent links maintain a minimum self-collision distance, implicitly guaranteed through $\Gamma(q,\dot q)$. The simulation runs for 6 seconds.

\textit{2) Joint Position and Velocity Limits \& External Obstacle Collision Avoidance:} We designed a simple scenario in which the robot is tasked with reaching the convergence point 
\(x^\ast = [\,0,\,-0.6,\,0.3]^\top\) while avoiding both a static and a dynamic obstacle along its path. 
The static obstacle is located at \([\,0.4,\,-0.3,\,0.4]^\top\), whereas the dynamic obstacle follows a 
trajectory \([\,0,\,-0.4,\,0.5+0.1\sin(2t_{\text{run}}))]^\top\) as a function of runtime \(t_{\text{run}}\). 
Both obstacles are modeled as spheres with a radius of \(5\,\text{cm}\). The simulation runs for 
4 seconds, during which the manipulator is required to maintain a clearance of at least 
\(5\,\text{cm}\) from the obstacle surfaces. As shown in Fig.~\ref{fig:sim-eca}, the incorporation 
of the external collision avoidance constraints ensures that collisions with the obstacles are prevented. The simulation runs for 4 seconds.

\textit{3) ALL Constraints:}
We designed a scenario where the robot would collide with a static spherical obstacle of radius 
\(5\,\text{cm}\), located at \([\,0,\,0.3,\,0.2\,]^\top\), on its way to the target position 
\([\,-0.1,\,0,\,0.3\,]^\top\). In addition, self-collision configurations arise if the SCA constraint is 
not enforced, as shown in the top two subfigures of Fig.~\ref{fig:sim-all}. By simultaneously 
applying joint-limit, self-collision, and external-collision constraints, our controller ensures 
safe and feasible motion throughout the trajectory, as illustrated in the bottom subfigure of 
Fig.~\ref{fig:sim-all}. The simulation runs for 6 seconds.

\textit{4) Comparison with CPIC:} We compare our method against the baseline CPIC controller with \textbf{three} sets of experiments. First, we conducted five experiments under the SCA setting,  where the robot’s initial joint configuration was randomly sampled and the target was fixed at 
\([\,0,\,0,\,0.3\,]^\top\). Similarly, five experiments were conducted under the ECA setting with one static obstacle;  both the obstacle position and the robot’s initial configuration were randomly sampled. It is worth noting that our VPP-TC enforces external-collision avoidance by considering the full geometry of all robot links, 
whereas in the version of CPIC available to us, only the last three links were taken into account.
Finally, five experiments were performed under the ALL-constraints setting, 
with the robot initialized at \([\,0,\,0,\,0,\,-1.5708,\,0,\,1.8675,\,0\,]^\top\), a static obstacle located randomly, and the target was fixed at \([\,-0.25,\,-0.35,\,0.5\,]^\top\). Each ALL trial was run for \(t\) seconds. For all experiments, we use three quantitative metrics to evaluate performance. \textit{(i) Control Loop Rate:} is defined as the average number of control steps executed per second,  with higher values indicating better performance.  In simulation, it is computed as the total number of simulated control steps divided by the actual wall-clock runtime on the computer. \textit{(ii) Path Length:} is the total distance traveled by the end-effector in simulation to accomplish the task. 
A shorter path length indicates a more efficient trajectory. \textit{(iii) Trajectory Jerkness:} reflects the smoothness of the executed trajectory. 
Let $x(t) \in \mathbb{R}^3$ be the end-effector trajectory over $t \in [0,T]$, with jerk $j(t) = \dddot{x}(t)$ 
and path length $L = \int_{0}^{T} \|\dot{x}(t)\|\,dt$. 
The normalized jerk is then defined as
\begin{equation}
  \mathrm{NJ} = \frac{1}{L^{2}T^{5}} \int_{0}^{T} \| \dddot{x}(t) \|^{2} \, dt,
  \label{eq:nj}
\end{equation}
where smaller values indicate smoother trajectories. As summarized in Table~\ref{tab:comp_cpic}, VPP-TC consistently achieves the lowest computation time, shortest path length to the task position, and significantly reduced trajectory jerkness vs. CPIC.

\begin{figure*}[!tbp]
    \centering
    \subfigure[Without SCA, the arm quickly triggers Reflex Mode due to self-collision and halts.\label{fig:real-sca-a}]{
        \includegraphics[width=0.315\textwidth]{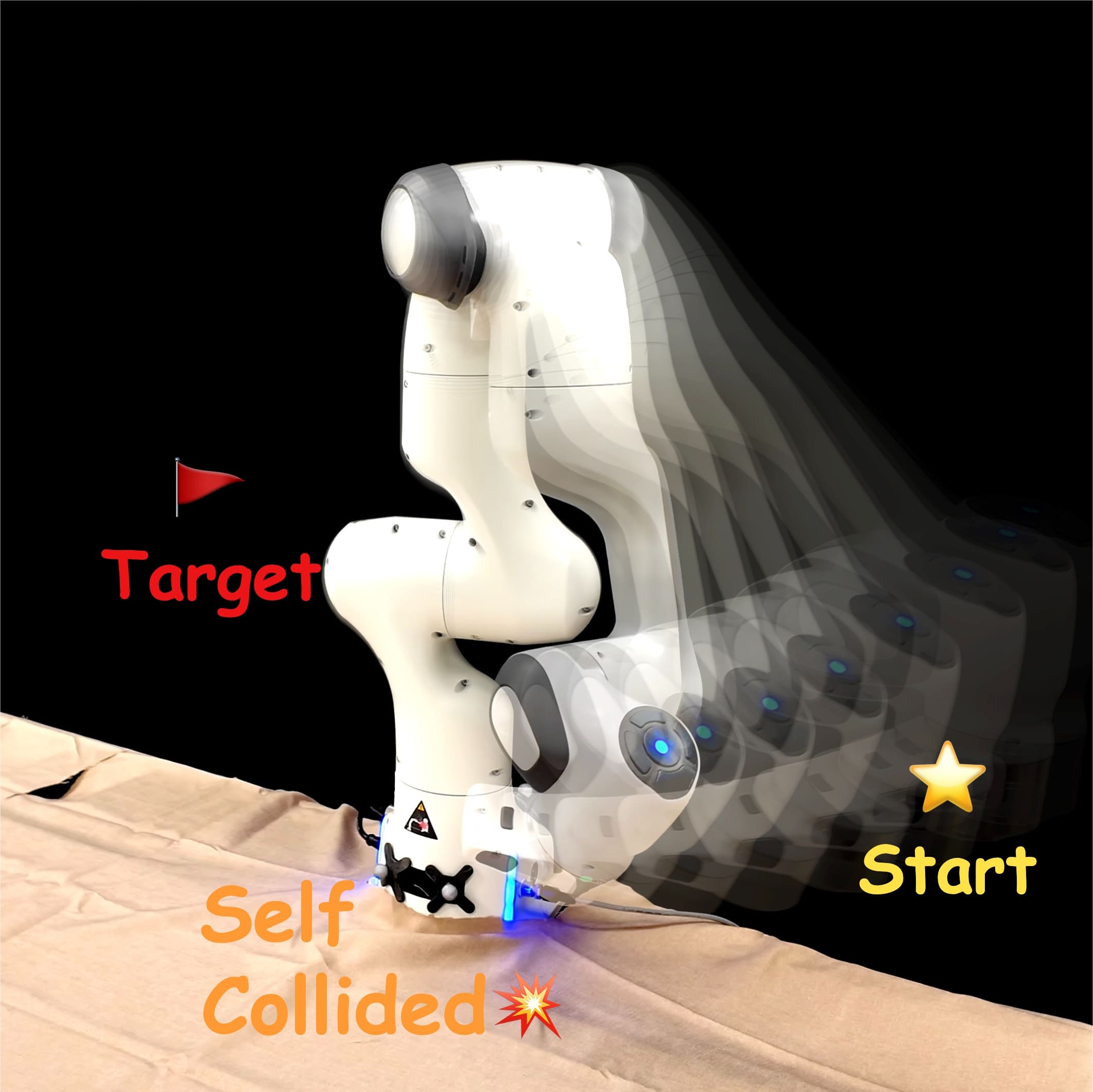}
    }\hfill
    \subfigure[With SCA enabled, the arm safely avoids self collision and reaches the target. \label{fig:real-sca-b}]{
        \includegraphics[width=0.315\textwidth]{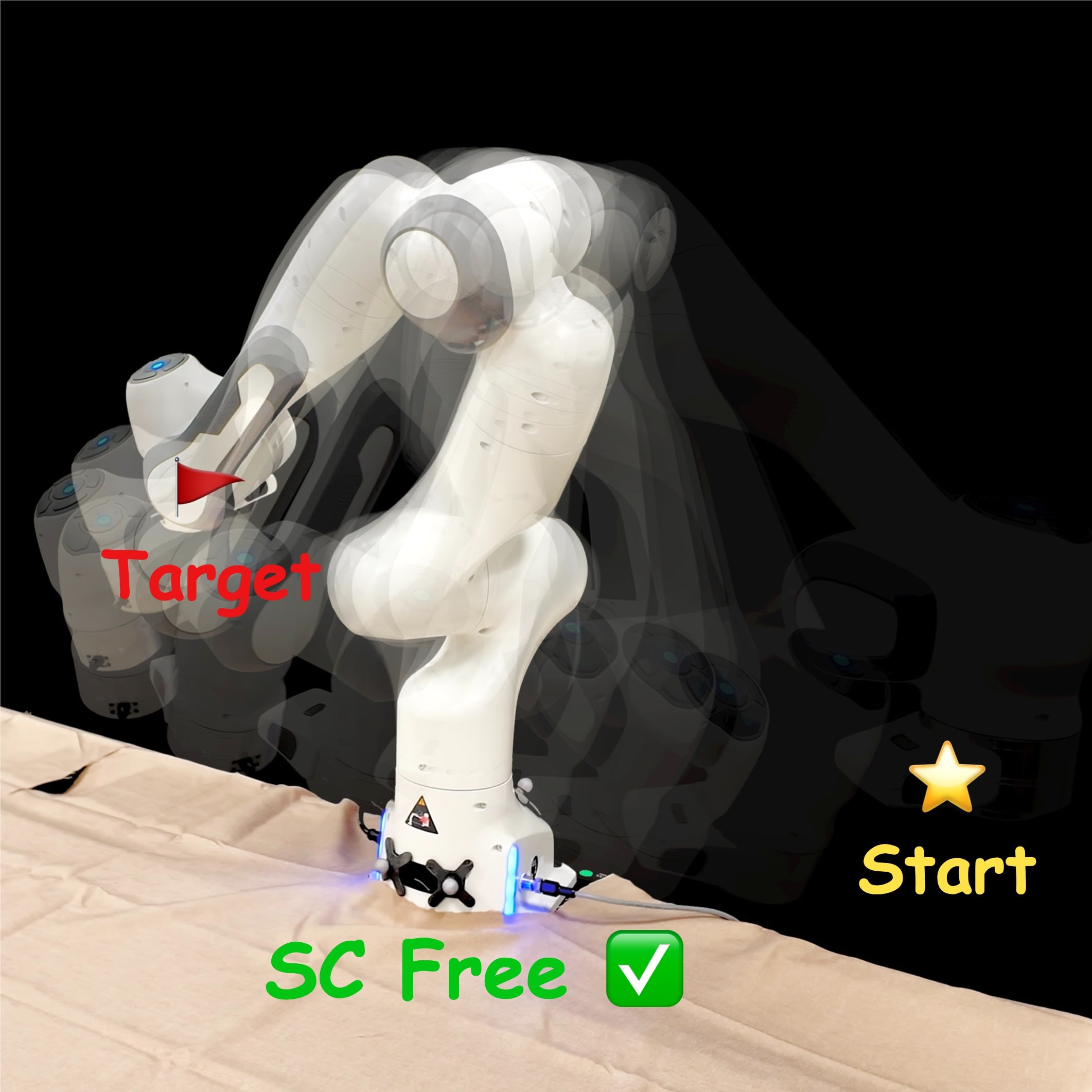}
    }\hfill
    \subfigure[With a UMI gripper, the arm cannot be pushed beyond the viability boundary.\label{fig:real-sca-c}]{
        \includegraphics[width=0.315\textwidth]{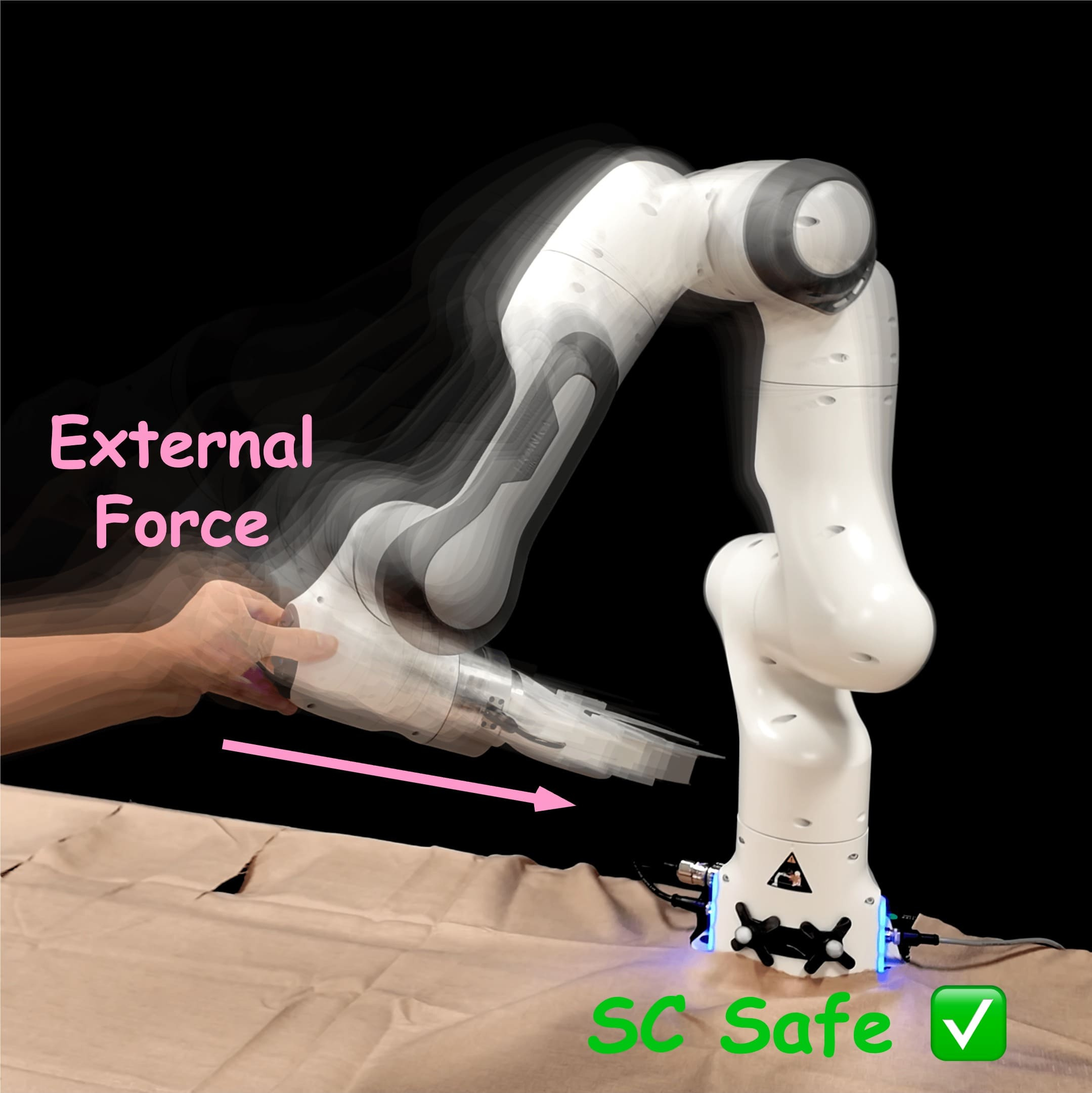}
    }
    \subfigure[Static obstacle with external push: the arm avoids obstacle and stops with a positive clearance.\label{fig:real-eca-a}]{
        \includegraphics[width=0.315\textwidth]{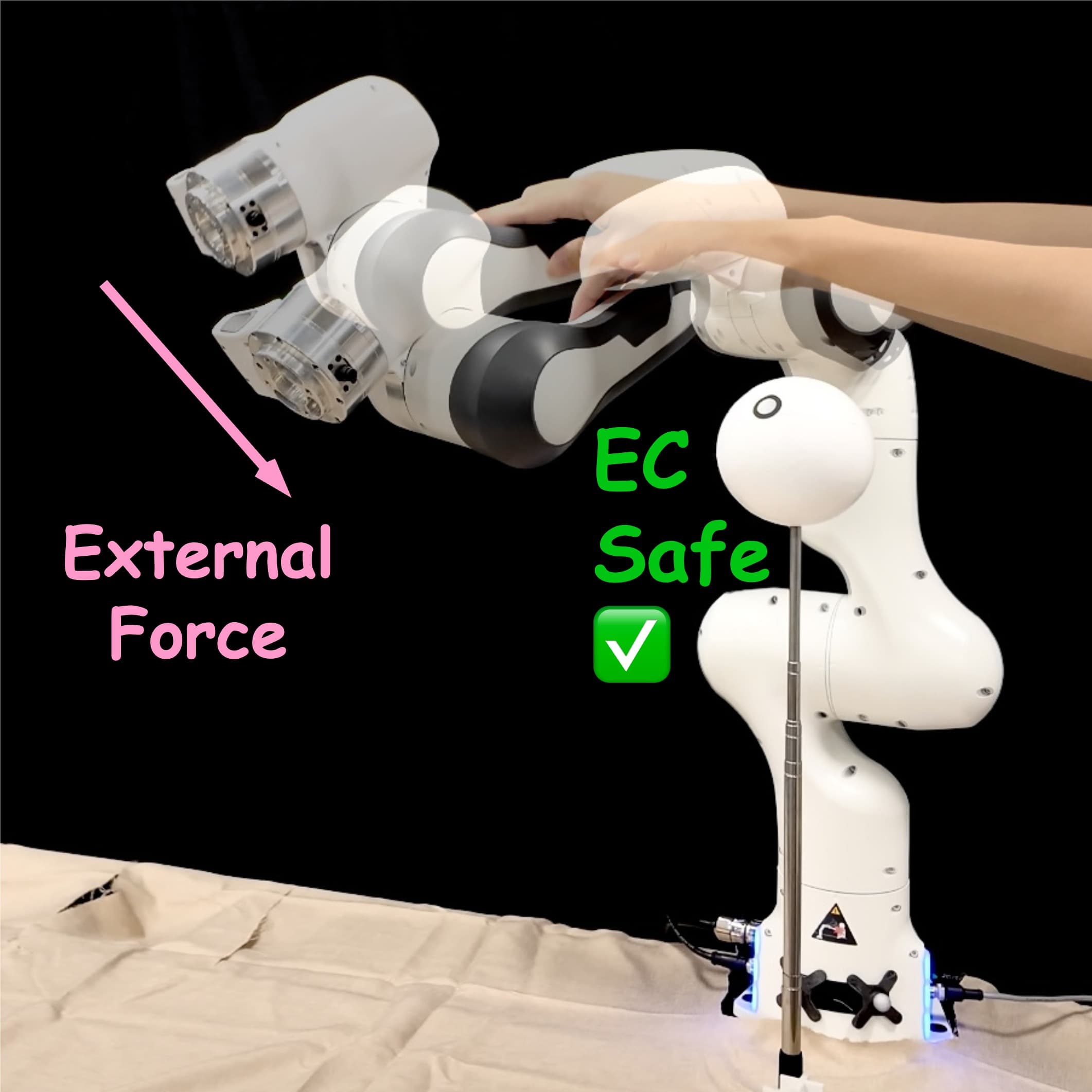}
    }\hfill
    \subfigure[Moving obstacle approaching: the arm yields and maintains a safe clearance from obstacle.\label{fig:real-eca-b}]{
        \includegraphics[width=0.315\textwidth]{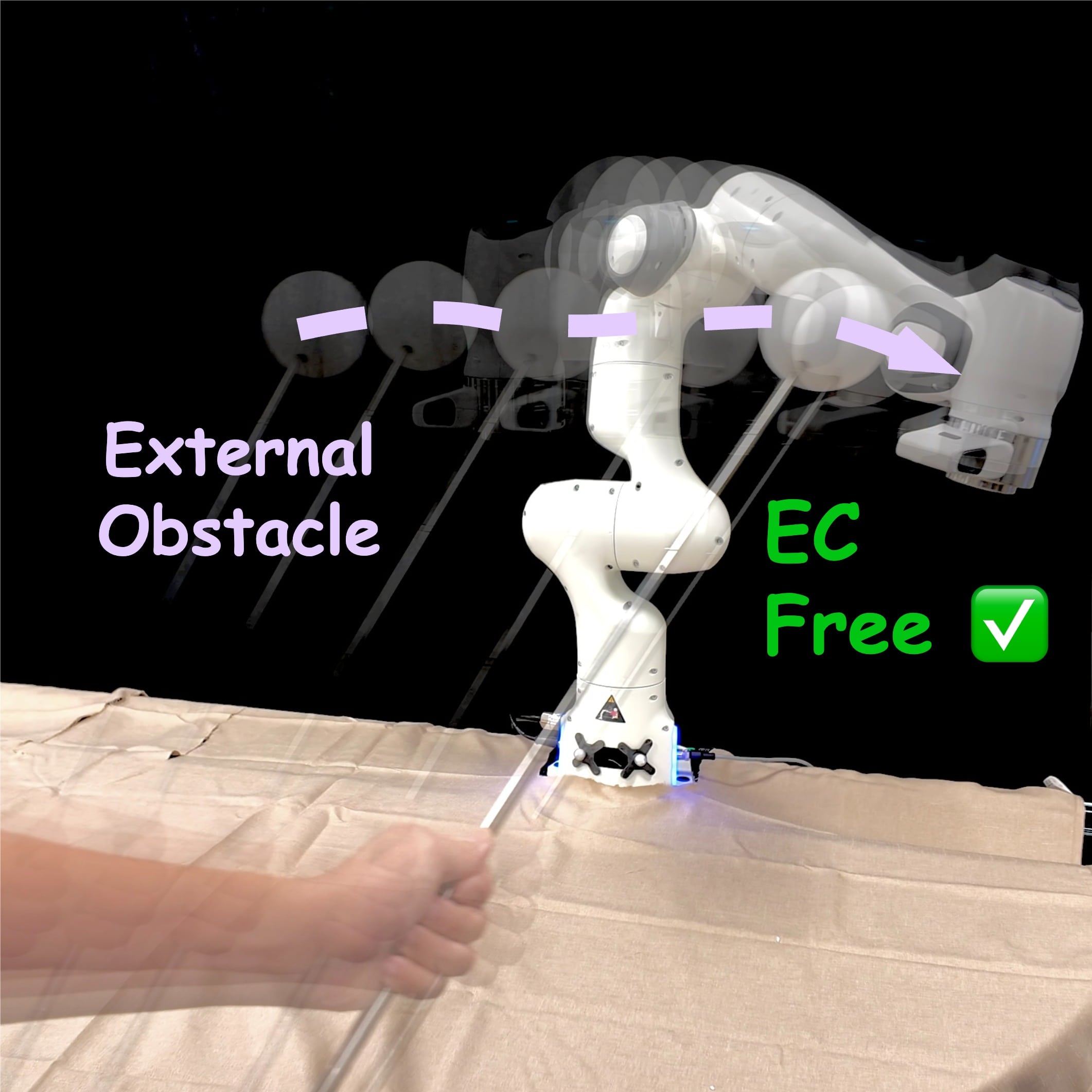}
    }\hfill
    \subfigure[Teleoperating near a lamp: positive clearance from obstacle preserved.\label{fig:real-eca-c}]{
    \includegraphics[width=0.315\textwidth]{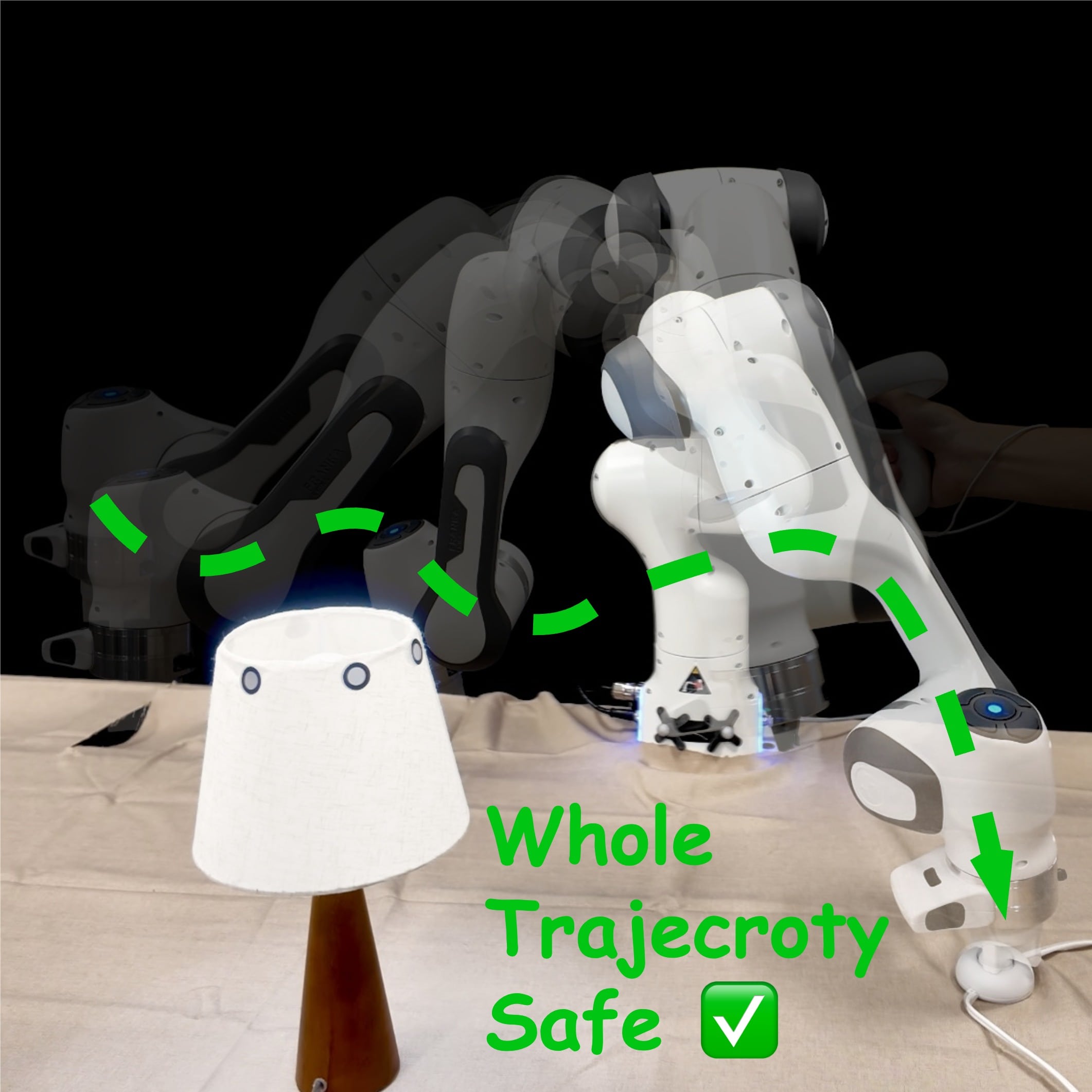}
    }
    \caption{Real-robot demonstrations of the proposed controller. Top row: self-collision avoidance (SCA). Bottom row: external-collision avoidance (ECA). Trajectory overlays show that activating the corresponding constraint prevents contact and maintains positive clearance while allowing the task to proceed.\label{fig:real_sca_eca}}
    \vspace{-10pt}
\end{figure*}

\subsection{Real-Robot Experiments}
In addition to simulation, we conducted three experimental scenarios on the 7-DoF Franka Panda to examine the controller’s effectiveness under real-world conditions: 
1) joint limits with self-collision avoidance, 
2) joint limits with external collision avoidance, and 
3) joint limits with both self-collision and external collision avoidance. 
Experiment~1) was performed on an Intel NUC 13 Pro running Ubuntu~20.04. 
Experiments~2) and~3) were conducted using both an Intel NUC 13 Pro and a workstation running Ubuntu~22.04, equipped with an Intel i7-11700K CPU and an NVIDIA RTX~3070 GPU. 
All real-robot experiments employed ROS Noetic for communication, where the workstation setup used \textit{RoboStack}~\cite{FischerRAM2021} to build the Noetic environment.

\textit{1) Joint Position and Velocity Limits \& Self-Collision Avoidance:} 
We performed three experiments in this setting. In the first two, the robot started from the joint configuration \([\,0.483,\,0.200,\,0.712,\,-2.240,\,-0.190,\,2.384,\,1.325\,]^\top\) at rest and was commanded to reach the target position \([\,0,\,-0.3,\,0.4\,]^\top\). Without the self-collision avoidance (SCA) constraint, the arm quickly entered Reflex Mode due to a self-collision warning and halted (Fig.~\ref{fig:real-sca-a}). With the SCA constraint enabled, the robot safely avoided self-collision and reached the target (Fig.~\ref{fig:real-sca-b}). In the third experiment, we attached a UMI gripper as the end-effector and retrained the neural network $\Gamma_{\text{umi}}(q,\dot q)$ to incorporate the new geometry, showing that the learned viability boundary generalizes across different robot geometries and continues to prevent unsafe states (Fig.~\ref{fig:real-sca-c}).
Moreover, the fact that the arm cannot be pushed beyond the boundary highlights that the controller enforces a normal constraint reaction and clamps the normal velocity to zero, while tangential damping dissipates motion. 
In this way, external input energy is either dissipated through damping or absorbed by the constraint reaction, and no net energy is injected into the environment, reflecting the task-space passivity of the controller.

\textit{2) Joint Position and Velocity Limits \& External Obstacle Avoidance:}
The positions of external obstacles were provided by an OptiTrack motion capture system. 
We first tested the case of a static spherical obstacle placed next to the robot. 
When external forces were applied to push the arm toward the obstacle from different joint configurations, 
the arm resisted the motion and stopped while maintaining a safe clearance before contact (Fig.~\ref{fig:real-eca-a}).
Similar to the self-collision scenario, as the arm approached the obstacle, it was unable to be pushed any further, indicating that the controller 
generated a reaction force along the obstacle normal while suppressing motion in that direction. 
At the same time, damping in the tangential directions dissipated the residual motion. 
Consequently, the external work was either absorbed by the constraint reaction or dissipated, and the system 
remained passive in task space without injecting net energy into the environment.
In a second experiment, we tested a moving spherical obstacle approaching the arm. 
The robot responded by actively moving away and maintained a safe clearance without collision (Fig.~\ref{fig:real-eca-b}). 
More dynamic and extreme scenarios with either multiple or high-speed moving obstacles are demonstrated in the 
accompanying video and on our project \href{https://vpp-tc.github.io/webpage/}{website}. 
Unlike the previous CPIC approach, which struggles with multiple or dynamic obstacles due to computational 
bottlenecks, our VPP-TC handles such cases effectively. 
The high update rate ($\sim$200\,Hz with two dynamic obstacles) enables true whole-body control with multiple obstacles, 
whereas CPIC in practice only considered the distance between the obstacle and the 7th link. 
Moreover, since the external-collision avoidance constraint~\eqref{eq:delsc} implicitly incorporates obstacle 
velocity through the $\Delta p$ term, VPP-TC can also robustly handle fast-moving obstacles.

\textit{3) ALL Constraints:}
In this experiment, the robot was teleoperated by specifying only coarse target positions rather than precise collision-free trajectories, 
and tasked with pressing the switch of a desk lamp to turn it on (Fig.~\ref{fig:real-eca-c}). 
Along the way to the switch, the planned path intersected with the lamp body (external obstacle) and also brought the arm close to the base link, 
creating a risk of self-collision. 
With VPP-TC, however, the robot avoided both external and self-collisions without any explicit motion planning, 
and safely reached the switch to accomplish the task (Fig.~\ref{fig:teleop_lamp}). 
This result highlights several key advantages of VPP-TC:  
\begin{itemize}
  \item It provides unified handling of joint limits, self-collision, and external collision constraints within the same framework.
  \item It enables safe execution of high-level teleoperation commands without requiring handcrafted obstacle-avoiding trajectories. 
  \item It ensures robust task-space passivity, so that safety is guaranteed under coarse or uncertain human inputs. 
\end{itemize}
These features underline the potential of VPP-TC as a versatile controller for safe human-in-the-loop manipulation.

\section{CONCLUSIONS AND FUTURE WORKS}
This paper proposed a novel viability-based control framework that enforces safety constraints for passive torque-controlled robots. We construct viable sets in the state space via a combination of data-driven learning and analytical derivation. A quadratic program then enforces the resulting torque bounds. We demonstrate the effectiveness and performance advantages of the proposed approach over a baseline through both simulation and hardware experiments. Our current control framework presumes an accurate dynamics model. In future work, we aim to accommodate bounded mismatches between the true and estimated dynamics and design a robust controller that guarantees viability in the presence of modeling errors.
\section*{Acknowledgement}
This work was supported by the National Science Foundation (NSF) Foundational Research in Robotics (FRR) program under NSF CAREER Award Grant No. FRR-2443721.
\bibliographystyle{IEEEtran}
\bibliography{ref.bib}

\appendix
\subsection{VPP-TC passivity analysis}\label{appdx-a}
We analyze the passivity of the controller generated by the simplest version of \eqref{eq:QP_framework} with just one constraint active,
\begin{equation}
    \min_{\tau_c}\ \frac{1}{2}\|J(q)^{-\top}\tau_c - F_c(x)\|_2^2,\ {\rm s.t.}\ a^\top \tau_c\leq b,
\end{equation}
where $a\in \mathbb{R}^n$ and $b \in \mathbb{R}$ are the parameters of the active affine constraint. We have the closed-form solution of the above QP as:
\begin{equation}
    \tau_c^* = J(q)^\top F_c - \frac{a^\top J(q)^\top F_c - b}{a^\top J(q)^\top J(q)a}J(q)^\top J(q)a.
\end{equation} We analyze the passivity in task-space, the equivalent task space force is expressed as:
\begin{equation}
\begin{aligned}
        F_c^* &= J(q)^{-\top}\tau_c\\ &= F_c - J(q)^{-\top}\frac{a^\top J(q)^\top F_c - b}{a^\top J(q)^\top J(q)a}J(q)^\top J(q)a\\
        &= (I - \frac{J(q)a(J(q)a)^\top}{(J(q)a)^\top J(q)a})F_c - \frac{bJ(q)a}{a^\top J(q)^\top J(q)a}\\
        &= \bar A F_c - \bar B,
\end{aligned}
\end{equation}
where $\bar A = I - \frac{J(q)a(J(q)a)^\top}{(J(q)a)^\top J(q)a}$, $\bar B = \frac{bJ(q)a}{a^\top J(q)^\top J(q)a}$. We define the storage function as $S = \frac{1}{2}\dot x^\top M \dot x + \lambda_1 P(x)$, where $P(x)$ is the potential energy. We can make the following derivation:
\begin{equation}
    \begin{aligned}
        \dot S- \dot x^\top f_{ext} =& \dot x^\top M \ddot x + \frac{1}{2}\dot x^\top \dot M \dot x + \lambda_1 \dot x^\top \nabla_x P\\
         =& \dot x (\bar A - I)G - \dot x \bar A D x - \dot x \bar B\\
        &+ \lambda_1 \dot x^\top (\nabla_xP + \bar A f(x)).
    \end{aligned}
\end{equation} We can conclude that for task-space state $(x, \dot x)$ satisfies $\dot x (\bar A - I)G - \dot x \bar A D x - \dot x \bar B + \lambda_1 \dot x^\top (\nabla_xP + \bar A f(x))\leq 0$, the robot is passive. Otherwise, passivity is lost and the robot becomes stiff and enforce the constraints.






	 
\end{document}